# Teaching Einsteinian Physics at Schools: Part 3, Review of Research Outcomes

**Tejinder Kaur[1], David Blair[1], John Moschilla[1], Warren Stannard[1] and Marjan Zadnik[1]**

[1]The University of Western Australia, 35 Stirling Highway, Crawley, WA 6009, Australia.

E-mail: tkaur868@gmail.com

**Abstract**
This paper reviews research results obtained from Einsteinian physics programs run by different instructors with Years 6, 9, 10 and 11 students using the models and analogies described in Parts 1 and 2. The research aimed to determine whether it is possible to teach Einsteinian physics and to measure the changes in students attitudes to physics engendered by introducing the modern concepts that underpin technology today. Results showed that students easily coped with the concepts of Einsteinian physics, and considered that they were not too young for the material presented. Importantly, in all groups, girls improved their attitude to physics considerably more than the boys, generally achieving near parity with the boys.

## 1. Introduction

Einsteinian physics underpins modern technology but is still not a part of most school science curricula. Many reasons have been suggested for the lack of focus on modern physics, including a) a belief that to understand modern physics students need to have a strong background in classical physics and b) that modern physics is conceptually too difficult.[1],[2] Numerous educators and popularisers have drawn attention to this problem, for example, as highlighted in a 'Minute Physics' video[3] which decries the historical nature of physics education. Recently physics education researchers from Australia, Norway and Scotland suggested that the cause of the problem is that "many people imagine that Einstein's theories require enormous mathematical skills".[4] A growing number of educationalists have been questioning these views.

Previous research shows that the concepts of quantum physics can be taught in the classroom.[5] Walwema et al. reported that students without any physics background have capabilities to understand the basic ideas behind Einsteinian physics.[6] The authors emphasise that "physics of the 20th century should be taught—not just to high school students—but to all STEM students". In the Einstein-First program, we also observed that students were excited to know more about Einsteinian physics for the understanding of the working of modern technology that relies on Einsteinian physics. Kraus mentions that high school students could understand the concepts of general relativity if we present them conceptually rather than mathematically.[7]

The models and analogies presented in Part 1 and Part 2 of this series have been tested with Year 6, Year 9, Year 10 and Year 11 students, aged from 11 to 16 years. These programs were run by senior physics and education research project leaders, doctoral and masters students. The Year 6 program was run by authors David Blair and Marjan Zadnik, Year 9 program was run by Tejinder Kaur, Year 10 program was run by John Moschilla and Warren Stannard and Year 11 program was run by Tejinder Kaur, David Blair, and Marjan Zadnik.

To measure the effectiveness of these programs, pre and post testing was undertaken to test both conceptual knowledge and attitudes to physics. The conceptual knowledge pre/post and attitudinal pre/post questionnaires were designed by the authors (see Appendix). The "conceptual pre/post–tests" were designed to assess students' conceptual knowledge of Einsteinian physics before and after the program. These tests had identical open-ended questions such as 'What is light?', 'What do you mean by the term gravity?' The 15 minute conceptual pre-test was given at the beginning of the program, while the "conceptual post-test" was given at the end of the program under identical conditions to the pre-test.

The "attitudinal pre/post-tests" were designed to assess changes in student attitudes to physics engendered by the program. Both questionnaires had identical questions based on Likert scale items such as 'I think physics is an interesting subject' and 'I enjoy learning new concepts and ideas'. The 10-minute attitudinal pre-test was given at the beginning and attitudinal post-test after the program, immediately after the conceptual pre/post-tests. Questions were partially selected from previously validated test questions [8] but it was found necessary to introduce new questions related to the content of the program. Both conceptual pre/post and attitudinal pre/post-tests were analysed and compared. Some of these test results are given below.

The programs were based on the materials presented in Parts 1 and 2. Each program was delivered using similar models and analogies but with the presentation style and language appropriate to the age group. Each 45-minutes lesson was divided into three roughly equal components:
  a) Power point presentation to explain Einsteinian Physics concepts visually, including many pictures, video clips, animations, and key words,
  b) Activities based on the models and analogies described in Parts 1 and 2.
  c) Work time: the last 15-minutes was used to complete worksheets and allow class discussion.

The primary goal of the research was to answer the following questions:
  a) Is it possible to teach Einsteinian physics to younger students,
  b) Does learning of Einsteinian physics improve student attitudes to physics.

In the next section, we review selected results from all of our programs. Detailed analysis of each program will be published elsewhere. The purpose of this review is to validate the material presented in Parts 1 and 2, and to summarise our evidence that teaching Einsteinian physics by using simple and interactive models and analogies has the power to motivate, enthuse and stimulate students between the ages of 11 and 16. The results presented are the results of short programs. These results are based on post-questionnaires taken soon after completion and therefore do not measure long-term effects which will require further studies.

## 2. Results

*2.1 Year 6: Eight-week program on Einsteinian physics*
Firstly, we will present attitudinal results obtained from Year 6 students who attended six lessons on Einsteinian physics. The results presented here were taken from data given in Pitts et al. (2014).[9] Two of the attitude questions asked by the authors were: "Was it interesting to find out about space and time and gravity?" and "Do you feel you are too young to understand Einstein's ideas?". The results obtained from these two questions are shown below in Figure 1(a) and Figure 1(b).

Figure 1(a) shows that 70% of students responded that the ideas of space, time and gravity were very interesting. Only one student identified the program as boring. Figure 1(b) shows responses to the "too young" question. Five individuals (~19%) from the class responded that they were too young to understand Einsteinian physics; most of the class (14 or 54%) responded that they were not too young to understand Einsteinian physics, and others (7 or ~27%) were undecided. Pitts et al. inferred that Year 6 students were conceivably ready to understand Einsteinian physics ideas.

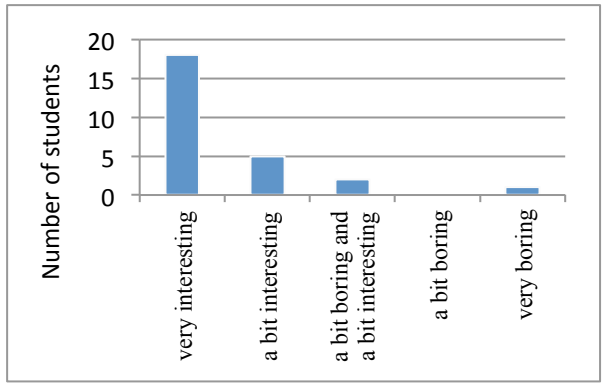
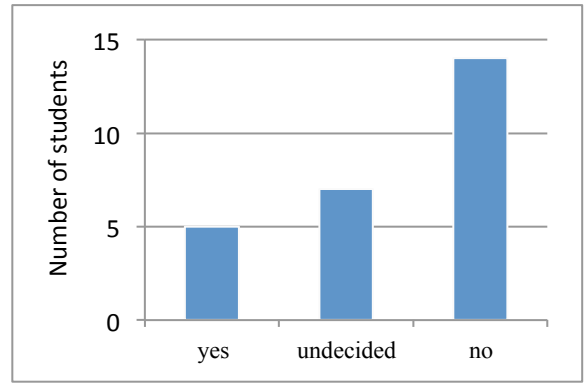

Figure 1(a). Year 6 student responses to the question: "Was it interesting to find out about space, time and gravity?" following a 6-lesson program on Einsteinian physics. 18 out of 26 students provided the most positive possible response. (Pitts et al., 2014).

Figure 1(b). Year 6 student responses to the question: "Do you feel you are too young to understand Einstein's ideas?" following a 6-lesson program on Einsteinian physics. 14 out of 26 students responded 'no' (Pitt et al., 2014).

## 2.2 Year 9: Ten-week program on Einsteinian physics

Next, we introduce the outcomes from Year 9 students' conceptual understanding based on pre- and post-questionnaires. These students were academically talented and went on a 10-week program on Einsteinian physics in 2014. Figure 2 indicates student understanding of Einsteinian physics before and after the program. It is clear from this figure that students' conceptual understanding improved dramatically following the program. Student pre-test scores were very low, with less than 10% of students scoring more than 50%. It is interesting that some of the students who had the lowest initial scores achieved results as high as the students with the highest pre-test scores. After the program, 53 out of 57 students scored 80% or above.

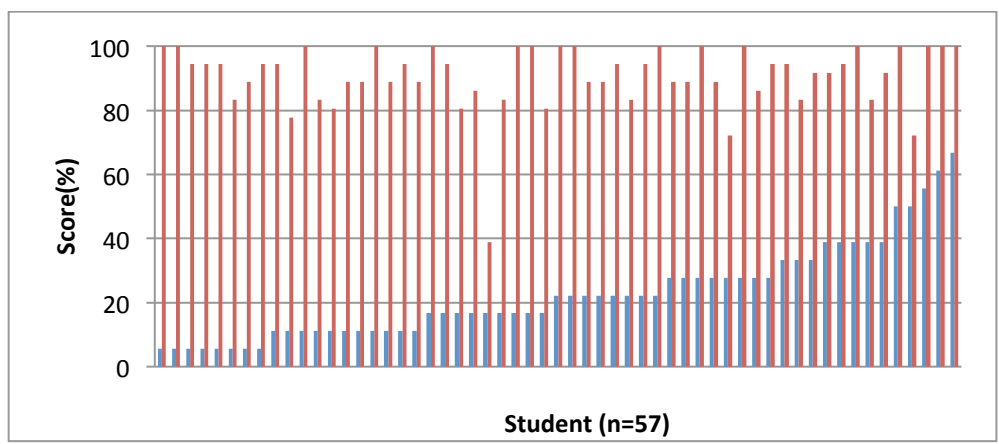

Figure 2. Pre-and post-test scores of 57 students conceptual understanding, arranged in order of ascending pre-test scores, of concepts of light, space and time following a twenty-lesson program on Einsteinian physics in 2014. All students demonstrated improvement in conceptual understanding.

We found that boys had a slightly greater interest in physics throughout the program, but that girls showed a significantly greater *increase* in interest. Figure 3 gives data showing that girls' interest toward physics improved from 50% to 80%, much greater than that for boys. The classroom teachers present during the program were also highly motivated by the activities used in the program.

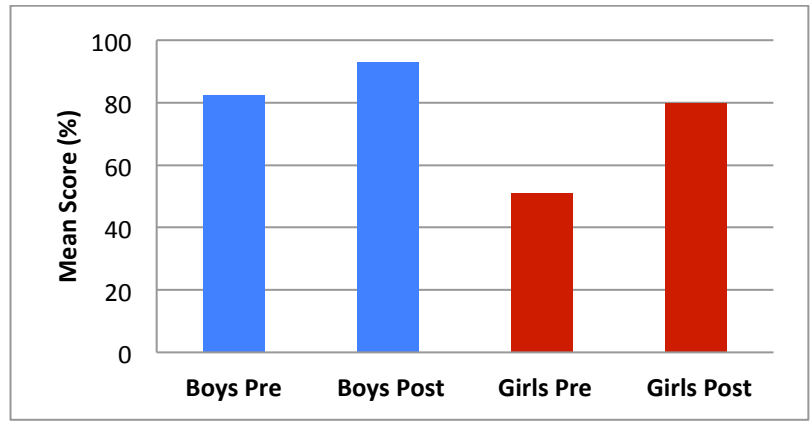

Figure 3. Student pre-and post-test responses to the question: "I think physics is an interesting subject" sorted by gender, following a twenty-lesson program on Einsteinian physics in 2014. Whilst both genders demonstrated improvement in attitude towards physics, the improvement in females scores was significantly greater than males.

## 2.3 Year 10: Four-week program on Einsteinian physics

A shorter four-lesson program was undertaken with Year 10 academically talented students. This program used some of the models and analogies described in Parts 1 and 2, while focusing on the Einsteinian concepts underlying the synthesis of gold in the universe, as well as its special properties. Figure 4 shows that students' pre-test scores were in the range 15-50%. Again, the students in the class improved significantly after the program. As shown in Figure 5, the improvement by girls again exceeded that of the boys. In this case, the girls performed lowest in the pre-test but exceeded the boys in the post-test.

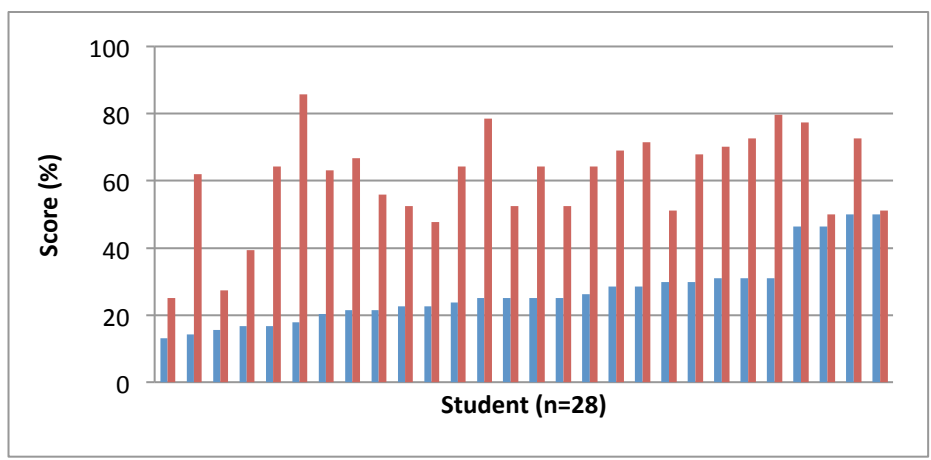

Figure 4. Pre-and post-test scores for Year 10 students conceptual understanding of Einsteinian physics concepts, arranged in order of ascending pre-test scores, following a four-lesson program on Einsteinian physics in 2015. All students demonstrated improvement in conceptual understanding of Einsteinian concepts.

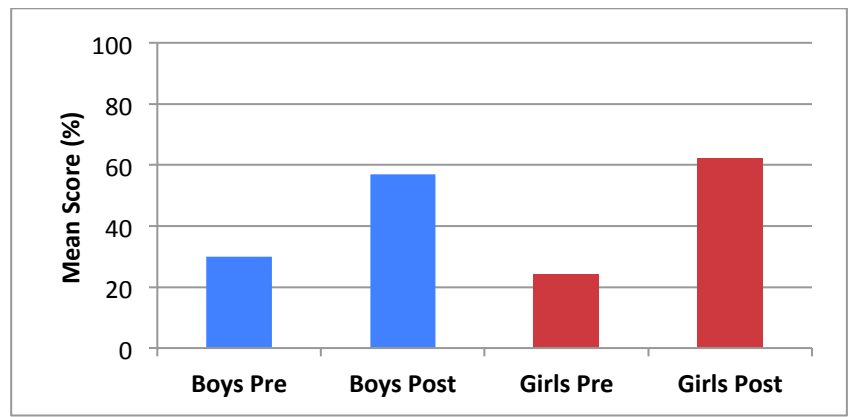

Figure 5. Averaged pre-and post-test scores for Year 10 students conceptual understanding of Einsteinian physics concepts, arranged in by gender, following a four-lesson program on Einsteinian physics in 2015. While both genders demonstrated significant improvement in their conceptual understanding of Einsteinian concepts, the improvement in female scores was significantly greater than the males.

*2.4 Year 11: One-day program on Einsteinian physics*

Lastly, we discuss a study of a one-day program with Year 11 students who were participants of the National Youth Science Forum, a national summer program. They were advanced students selected from different states of Australia. Figure 6 presents results from conceptual understanding questionnaires. These results show a similar trend to the results obtained with younger students. Students' conceptual understanding before the program was low, although a greater fraction of the class scored near 50%. Following the program, all but two students improved their test scores, with an average improvement factor 2.1. However, the large variation in score improvement relative those found in the longer programs may indicate that a single-day program is insufficient for consolidating conceptual understanding of Einsteinian physics.

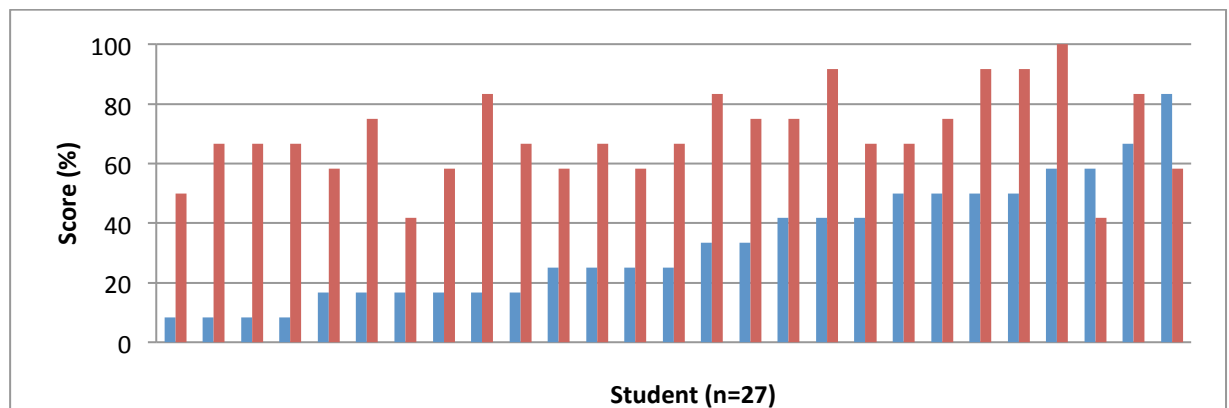

Figure 6. Pre-and post-test scores for 27 students conceptual understanding of Einsteinian physics concepts, arranged in order of ascending pre-test scores, following a one-day program on Einsteinian physics in 2014. All students but 2 demonstrated improvement in conceptual understanding. The mean test score result increased by a factor of 2.1 between the pre and post test.

## 3. Conclusion

It is interesting that similar models and analogies can be effective teaching tools over the breadth of school education from the ages of 11 to 16. The pre-test conceptual knowledge results showed that over the age range students had very little knowledge of fundamental concepts behind Einsteinian physics. Even out of nationally selected 16-years old students who participated in the National Youth Science Forum program, only 25% of students achieved pre-test scores of about 50%, while for 11-year-olds, 10% scored to this level.

Overall, it is very clear that Einstein-First programs were well within the ability of students over all age groups. All classes demonstrated substantially improved conceptual understanding of Einsteinian physics. The largest program undertaken with Year 9 students showed the most remarkable improvement with average class scores increasing from 23% to 91%. It was very clear in all programs that the learning of Einsteinian physics enhanced student attitudes and interest in learning Einsteinian physics.

The authors were not surprised to learn how well Einsteinian physics was accepted by students, but the results on gender were unexpected. The greater improvement factor of girls over boys has been replicated in all our studies. The girls enter with lower attitude and knowledge scores, and leave the program achieving near parity. We attribute this partly to the fact that students are aware that the current curriculum is based on old fashioned concepts, and partly because the girls may be more appreciative of conceptual ideas and active group based learning rather than conventional approaches. The result indicates that the introduction of Einsteinian physics, and the introduction of more conceptual learning may have a strong benefit in encouraging gender balance in STEM education and career choices. The research outcomes reviewed here are the results of programs that were piloting and exploring the possibility of introducing Einsteinian physics at an early age. Our testing has not evaluated long-term retention of conceptual understanding, nor the effects of a proper learning progression that would carry through several years and would integrate Einsteinian physics with Newtonian approximations. We believe that the results presented make a strong case for further research aimed at reconstructing school science within Einsteinian physics paradigm.

## 4. Acknowledgements

This research was supported by the Australian Research Council, the Gravity Discovery Centre and the Graham Polly Farmer Foundation. We thank Grady Venville and David Treagust for their enthusiastic support, and teachers Warwick Mathew, Dana Perks and Richard Meagher for their enthusiastic support for our programs.

**Appendix**

Below we give some typical questions used to assess knowledge and attitudes to physics before and after the programs discussed in this paper. We asked open ended questions in conceptual pre/post tests and attitudinal questionnaire was based on Likert scale items.

**Conceptual questionnaire**
1. Can parallel lines meet?
2. Can the sum of the angles in a triangle be different from 180 degrees?
3. What do you mean by the term "Light"?
4. Does space have a shape? Circle Yes or No. How could you measure the shape of space?
5. If you weighed an object on a supersensitive balance, would the balance register a different weight if you heated the object up?
6. How could you tell if a ruler is straight?
7. In the absence of air resistance, (like in a huge vacuum tank or on the moon) if we drop a hammer and a feather, which one of them will touch the ground first?
8. List the names of at least four types of electromagnetic radiation.
9. A person claims on Facebook that he has made a perfect microscope that is so accurate that the exact position of an atom can be measured. Could this claim be plausible?

**Attitudinal questionnaire**
1. I think physics is an interesting subject
2. I prefer to learn physics through activities
3. I enjoy learning new concepts and ideas
4. I enjoy trying things out at home and/or telling my family about school science activities.
5. I think doing activities helps me understand and remember new ideas much better than if it is just from books and lessons
6. The things that Einstein discovered are important for modern technology
7. I like doing calculations
8. Understanding scientific ideas are more important than memorizing facts
9. I enjoy science excursions and would like to have more of them.